\documentclass[letterpaper, 10pt, conference]{ieeeconf}
\usepackage{amsmath,amssymb,textcomp}
\usepackage{graphicx,mathrsfs,cuted}
\usepackage{multirow}
\usepackage{xcolor}

\IEEEoverridecommandlockouts                             

\newtheorem{theorem}{Theorem}

\newtheorem{remark}[theorem]{Remark}

\begin{document}

\title{Markovian Embeddings of Non-Markovian Quantum Systems: Coupled Stochastic and Quantum Master Equations for Non-Markovian Quantum Systems}

\author{Hendra I. Nurdin\thanks{H. I. Nurdin is with the School of Electrical Engineering and 
Telecommunications,  UNSW Australia,  Sydney NSW 2052, Australia (\texttt{email: h.nurdin@unsw.edu.au})}} 


\maketitle

\begin{abstract}
Quantum Markov models are employed ubiquitously in quantum physics and in quantum information theory due to their relative simplicity and analytical tractability. In particular, these models are known to give accurate approximations for a wide range of quantum optical and mesoscopic systems. However, in general, the validity of the Markov approximation entails assumptions regarding properties of the system of interest and its environment, which may not be satisfied or accurate in arbitrary physical systems. Therefore, developing useful modelling tools for general non-Markovian quantum systems for which the Markov approximation is inappropriate or deficient is an undertaking of significant importance. This work considers non-Markovian principal quantum systems that can be  embedded in a larger Markovian quantum system with one or more compound baths consisting of an auxiliary quantum system and a quantum white noise field, and derives a set of coupled stochastic and quantum master equations for embedded non-Markovian quantum systems.  The case of a purely Hamiltonian coupling between the principal and auxiliary systems as a closed system without coupling to white noises is included as a special case. The results are expected to be of interest for (open-loop and feedback) control of continuous-time non-Markovian systems and studying reduced models for numerical simulation of such systems.  They may also shed more light on the general structure of continuous-time non-Markovian quantum systems. \emph{[This work was published in Proceedings of the 62nd IEEE Conference on Decision and Control (Singapore, Dec. 12-15)  pp. 5939-5944 (2023)]}
\end{abstract}

\section{Introduction}

Quantum Markov models are an important class of models that is widely employed in various areas of quantum physics and quantum information for modelling quantum systems and quantum noise. For instance, quantum Markov models describe various quantum optical, optomechanical and superconducting systems, see, e.g,. \cite{GZ04,WM10,NY17} and are used to model quantum noise in quantum devices such as quantum computers \cite[Chapter 8]{NC10}. These models are popular because of their relative tractability for analysis. However, in the quantum context, different notions or viewpoints of Markovianity have been adopted, which are in general not equivalent. The choice of the definition of Markovianity that has been adopted is motivated by the application of interest. These different viewpoints will  be briefly discussed  by using the definition of classical Markov stochastic processes as a starting point of discussion. 

In the classical setting, a continuous-time Markov process $\{X_t\}$ on the nonnegative real line $t \geq 0$ taking on values in $\mathbb{R}^n$ can be equivalently defined as a process satisfying the property that
\begin{equation}
\mathbb{E}\left[g(X_t) | \mathcal{F}_{0:s}\right]= \mathbb{E}\left[g(X_t) | \mathcal{F}_s\right]\;\hbox{a.s.} \label{eq:Markov}
\end{equation}
for $ 0 \leq s \leq t$ and any bounded Borel measurable real function $g:\mathbb{R}^n \rightarrow \mathbb{R}$, where $\mathcal{F}_{0:s} = \sigma(X_{0:s})$ denotes the $\sigma$-algebra generated by the Markov process up to time $s$, $X_{0:s}=\{X_{\tau}, 0\leq \tau \leq s\}$, and $\mathcal{F}_s = \mathcal{F}_{s:s}$. If the Markov process has a  conditional probability density function (PDF) then one can write:
\begin{align*}
\mathbb{E}\left[g(X_t) | \mathcal{F}_{0:s}\right](X_{0:s})
&=\int_{\mathbb{R}^n} g(x) p_{X_t \mid \mathcal{F}_{s}}(x| X_{s}) dx, 
\end{align*}
where $p_{X_t \mid \mathcal{F}_{s}}(\cdot \mid X_{s})$ is the conditional PDF of $X_t$ given $X_{s}$. For a continuous-time Markov process, the evolution of the PDF $p_{X_t}(\cdot) = \mathbb{E}[p_{X_t | \mathcal{F}_{s}}(\cdot \mid X_{s})]$ from an initial distribution $p_{X_0}(\cdot)$ is given by a partial differential equation known as the Fokker-Planck equation.

In the quantum setting, an open quantum system (a quantum system that is interacting with some external environment) that is not being observed (monitored) can be described by its  reduced density operator by taking the partial trace over its environment.  Throughout the paper, an open quantum system of interest will be referred to as the {\em principal system}. Under certain  assumptions and approximations, there is class of quantum systems whose reduced density operator solves the Gorini-Kossakowski-Sudarshan-Lindblad (GKSL) quantum master equation:
\begin{equation}
\dot{\rho}_t = i[\rho_t,H] + L \rho_t L^{\dag} - \frac{1}{2} \{\rho_t,L^{\dag}L\}, \label{eq:GKSL}
\end{equation}
where $H$ is the Hamiltonian  of the principal system (a self-adjoint/hermitian operator) and $L$ is an operator of the principal (not necessarily hermitian) describing some coupling  to the  environment. The GKSL equation is the quantum analogue of the Fokker-Planck equation for a classical continuous-time Markov process. However, although the  PDF of a Markov process $\{X_t\}$ satisfies the Fokker-Planck equation, the process is not defined by this PDE but by the property \eqref{eq:Markov}. Indeed, non-Markov processes can be constructed whose PDF verifies the Fokker-Planck equation, see, e.g., \cite{McCaul12}. 

A quantum version of Markov processes that is based on a non-commutative generalization of the Markov property \eqref{eq:Markov} was introduced by Accardi, Frigerio and Lewis (AFL) \cite{AFL82} in an operator algebraic framework; see \cite{Nurd20} for an overview. A concrete realization of the AFL quantum stochastic process is given by an open quantum system that is coupled to a quantum white noise process as a memoryless environment. The joint unitary evolution of the open quantum system and the quantum white noise given by the solution to a Hudson-Parthasarathy quantum stochastic differential equation; see \cite{Nurd20} for a discussion. However, there will be non-Markovian quantum processes satisfying the GKSL equation, analogous to the classical setting. One such process is described in \cite[Appendix, Example A.1]{AFL82}, due to G. Lindblad; see also \cite{GN17}. 

In many applications of quantum information theory, such as quantum computing, the main object of interest is the reduced density operator of the system as this must be manipulated and controlled for various tasks. In this setting, a quantum Markov process is viewed as a process satisfying the GKSL equation \eqref{eq:GKSL} although such a process is not necessarily a quantum Markov process in the sense of AFL. In the AFL theory, the environment of a quantum system is included in the definition of a quantum stochastic process while in the GKSL-based view the environment is traced out. 


Classical non-Markovian stochastic processes are often treated by introducing additional variables so that the process becomes embedded in a larger Markov process. This allows the theoretical machinery and analytical tools of Markov processes to be applied to study a non-Markov process of interest. This strategy has also been adopted in the quantum setting  by introducing additional physical degrees of freedom that are traced out, leaving a quantum system with non-Markovian dynamics whose reduced density operator need {\em not} satisfy the GKSL equation. In particular, a quantum system coupled to a Gaussian bosonic noise with a non-flat spectral density, which means that the noise is colored (not a quantum white noise process), can be approximated by replacing the noise with a {\em compound noise model} given by a  finite collection of (fictitious) auxiliary single mode resonators that are driven by  quantum white noise processes \cite{Imamoglu94,DBG01,Mascherpa20}. An adaptation of the approach to colored fermionic noise is given in \cite{CAG19}. 


Another type of Markovian embedding in the literature  different from \cite{Imamoglu94,DBG01,Mascherpa20}, takes as the compound noise the continuous-mode output of a quantum input-output system \cite{GZ04,CKS17} that is driven by quantum white noise fields. The principal system is another quantum input-output system that is coupled to this compound noise via a cascade connection where the output from the compound noise drives the input of the principal system. This has been used to model quantum input-output systems that are driven by a bosonic quantum noise process in various non-Gaussian quantum states \cite{GJN11,GJNC12,GJN14}.  
The quantum Markovian embedding of a non-Markovian quantum system is the theoretical framework adopted in this work. We also note the use of non-Markovian embeddings to derive quantum filtering and master equations for systems driven by quantum white noise in a non-Gaussian state \cite{GJN13,SZX16}, which is distinct from the topic of this paper. 

This paper derives coupled stochastic and quantum master equations that is expected to be of interest for studying (open-loop and feedback) control of continuous-time non-Markovian systems and reduced models for numerical simulation of such systems. They may also shed additional light on the general structure of the evolution of continuous-time non-Markovian quantum systems. 

\section{Mathematical preliminaries}
\label{sec:background}
\textbf{Notation.} $X^{\top}$ denotes the transpose of a matrix $X$, $X^{\dag}$ denotes the adjoint of a Hilbert space operator $X$ and if $X=[X_{jk}]$ is a matrix of operators then $X^{\dag}$ is the conjugate transpose of $X$, $X^{\dag} =[X_{kj}^{\dag}]$. $I_n$ will denote an $n \times n$ identity matrix and $I$ can denote either an identity matrix (whose dimension can be inferred from the context), an identity map or an identity operator. $\mathrm{Tr}$ denotes the trace of a matrix or an operator and  $\mathrm{Im}(X)$ is the elementwise imaginary part of a matrix $X$. For a signal  (function of time) $Y$, $Y_{0:t}=\{Y_{\tau}\}_{0 \leq \tau \leq t}$. If $X$ is an operator on the composite Hilbert space $\mathfrak{h}_1 \otimes \mathfrak{h}_2$ then $\mathrm{Tr}_{\mathfrak{h}_j}(X)$ denotes the partial trace of $X$ by tracing out the Hilbert space $\mathfrak{h}_j$ ($j=1,2$). Also, $\mathbb{E}[\cdot]$ denotes the classical expectation operator. 

In the following, a brief overview of the notion of a quantum input-output (I/O) system will be given, see \cite{GZ04,GJ08,CKS17} for further details. To focus on the main ideas, only quantum I/O systems coupled only to a single traveling field will be discussed. Consider an open quantum system at a fixed location that interacts with a unidirectional quantum field which is traveling towards the system.  Under some physical assumptions and approximations, in a large class of physical scenarios of interest the unitary propagator $U_t$ on the system and the field, when the latter is in the vacuum state $|0_{\mathrm{f}}\rangle$, is given by a  Hudson-Parthasarathy quantum stochastic differential equation (QSDE) \cite{HP84}:
\begin{align}
dU_t &= (-(iH(t)+(1/2)L(t)^{\dag}L(t))dt + dB^{\dag}_t L(t) \notag \\
&\quad - L(t)^{\dag}S(t) dB_t + (S(t)-I) d\Lambda_t)U_t,\,U_0=I. \label{eq:HP-QSDE}  \end{align}
Here $B_t$, $B^{\dag}_t$ and $\Lambda_t$ are the annihilation, creation and gauge processes of the traveling field, $H(t)$ is the principal system Hamiltonian $L(t)$ is the coupling operator of the principal to the field creation operator, and $S(t)$ is a unitary operator ($S(t)^{\dag}S(t)=S(t)S(t)^{\dag}=I$ for all $t \geq 0$) representing the coupling of the system to the gauge process, all of which  can be time-dependent in general. The processes $B_t$, $B^{\dag}_t$ and $\Lambda_t$ are referred to as {\em fundamental processes}.  Note that there is no loss of generality in taking the vacuum state as more general Gaussian states of a quantum field can be realized through a combination of independent quantum fields in the vacuum state by the generalized Araki-Woods representation \cite{GJN10} \cite[\S 2.7.1.1]{NY17}. 

The evolution of a principal system operator $X$, in the Heisenberg picture with respect to the propagator \eqref{eq:HP-QSDE} is given by $j_t(X)$, where $j_t(X)=U_t^{\dag} X U_t$. It is given  by the QSDE:
\begin{align}
dj_t(X) &=\mathcal{L}_{j_t(L),j_t(H)}(j_t(X)) dt + dB_{t}^{\dag}j_t(S)[j_t(X),j_t(L)] + \notag \\
&\qquad [j_t(L^{\dag}),j_t(X)]dB_{t} \notag \\
&\quad +\mathrm{tr}(j_t(S^{\dag})j_t(X) j_t(S)-j_t(X))d\Lambda_{t}, \label{eq:X-evolution} 
\end{align}
where $\mathcal{L}_{Y,Z}(X)$ is a map defined by:
$$
\mathcal{L}_{Y,Z}(X) = i[Z,X] + (1/2) \left( Y^{\dag}[X,Y] + [Y^{\dag},X]Y  \right).
$$

Due to the interaction with the system, the fundamental processes that impinge upon the system at time $t$, considered as an {\em input field} to the system, undergo an instantaneous transformation according to $M_{{\rm o},t}=U_t^{\dag} M_t U_t$, where $M_t$ can be any of the fundamental processes or linear combinations thereof, producing {\em an output field}. Let $Y^{Q}_t = B_{t} + B^{\dag}_t$ and   $Y^{P}_t = -iB_t + i B^{\dag}_t$ be the amplitude and phase quadratures of $B_t$, respectively. Then $Y^{Q}_t$, $Y^{P}_t$ and $\Lambda_t$ are transformed to the  output field processes $Y^{Q}_{{o},t}$, $W^{P}_{{o},t}$ and $\Lambda_{{ o},t}$  given by the QSDE: 
\begin{align*}
dY^{Q}_{{ o},t} &=j_t(L(t)+ L(t)^{\dag})dt + j_t(S(t)) dB_{t}  +  j_t(S(t)^{\dag})dB^{\dag}_{t} \\
dY^{P}_{{o},t} &=j_t(-iL(t)+ i L(t)^{\dag})dt -i j_t(S(t)) dB_{t} \\
&\quad + i  j_t(S(t)^{\dag})dB^{\dag}_{t} \\
d\Lambda_{{o},t}  &=  j_t(L^{\dag})j_t(L(t)) dt  + j_t(S(t)^{\dag})j_t(L) dB_{t}^{\dag} \\
&\quad + j_t(L(t)^{\dag})j_t(S(t)) dB_{t} +d\Lambda_{t}.     
\end{align*}

Since the input and output fields are both unidirectional traveling fields  that are not constrained by boundary conditions, they contain a continuum of frequencies from $-\infty$ to $\infty$. Hence they are also referred to a {\em continuous-mode (input/output) quantum fields}. Measurements of $Y^{Q}_{{o}}$ and $Y^{P}_{{ o}}$ correspond to homodyne measurements of the amplitude and phase quadratures of the field, respectively, while  measurement of $\Lambda_{{o}}$ is a photon counting measurement. It is common and often useful to consider the Schr\"{o}dinger picture in which the system-field state evolves in time by applying the unitary $U_t$ to the initial state. This gives the state 
$$
\tau_t = U_t(\rho_0 \otimes |0_{\rm f}\rangle\langle 0_{\rm f}|) U_t^\dagger
$$
at time $t \geq 0$, where $\rho_0$ is the initial state of the system while, as before, $|0_{\rm f}\rangle$ is the vacuum state of the field. The reduced system state is $\rho_t= {\rm Tr}_{\mathfrak{h}_{\mathrm{f}}} (\tau_t)$, where $\mathfrak{h}_{\mathrm{f}}$ denotes the Hilbert space of the traveling quantum field, and it satisfies \eqref{eq:GKSL} with the substitutions $H(t) \rightarrow H$ and $L(t) \rightarrow L$.

\section{Markovian embedding of non-Markovian quantum systems}
\label{sec:Markov-embedding}
\subsection{Direct coupling embedding}

\begin{figure}
\centering
\includegraphics[scale=0.5]{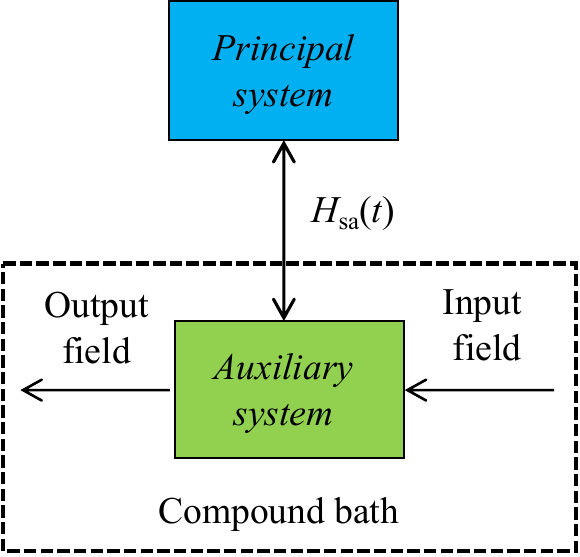}
\caption{Embedding by direct coupling. A principal quantum is directly coupled to the auxiliary quantum system in a compound bath consisting of  the auxiliary coupled to a single traveling quantum  field.}  \label{fig:direct} 
\end{figure}

The Markovian embedding in \cite{Imamoglu94,DBG01,Mascherpa20} involve only direct coupling between the principal system and the fictitious auxiliary degrees of freedom, with no direct coupling between  the principal and the quantum fields. For simplicity of discussion and to focus on the key ideas, consider only a single auxiliary degree of freedom coupled to a single quantum field in the vacuum state; see Fig.~\ref{fig:direct}. The principal has Hilbert space $\mathfrak{h}_{\rm s}$ while the auxiliary system has Hilbert space $\mathfrak{h}_{\rm a}$. The principal and auxiliary has the Hamiltonian $H_{\rm s}(t)$ and $H_{\rm a}(t)$, that can be time-dependent. They are coupled through an interaction Hamlltonian $H_{\rm sa}(t)$  that acts on both sub-systems. The auxiliary is in turn coupled to a quantum field in the vacuum state through a coupling operator $L_{\rm a}(t)$. The joint unitary propagator the principal + auxiliary + quantum field is given by the solution of the QSDE (for simplicity, it does not include a coupling of the auxiliary to the gauge process  $\Lambda_t$):
\begin{eqnarray*}
\lefteqn{dU_t}\\ 
&=& \left(-(i (H_{\rm s}(t) + H_{\rm a}(t) + H_{\rm sa}(t) + (1/2)L_{\rm a}(t)^{\dag}L_{\rm a}(t))dt \right.\\
&&\quad \left. + dB_t^{\dagger} L_{\rm a}(t)- L_{\rm a}^{\dag}(t)dB_t\right) U_t,\;U_0=I.
\end{eqnarray*}

\subsection{Cascaded embedding}
\begin{figure}
\centering
\includegraphics[scale=0.5]{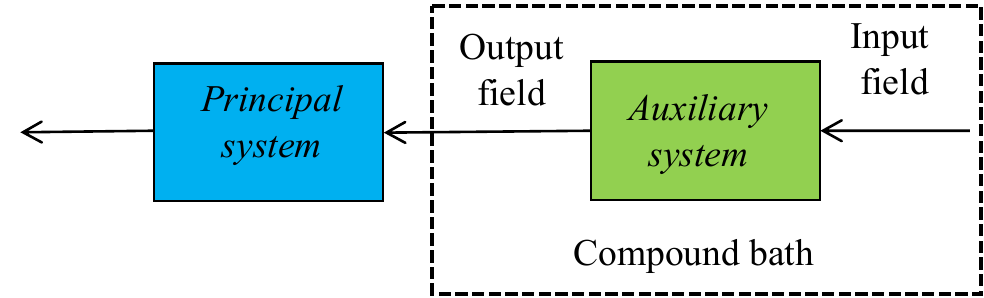}
\caption{Markovian embedding via cascading. A principal quantum I/O system is cascaded to a compound bath consisting of an auxiliary quantum I/O system that is coupled to a single traveling quantum field. The output field of the compound bath becomes the input to  the principal system.} \label{fig:cascaded} 
\end{figure}

Another type of non-Markovian dynamics arises by cascading a quantum I/O system to another quantum I/O system driven by a single quantum field as the compound bath, as shown in Fig.~\ref{fig:cascaded}. The principal quantum system is an I/O system with Hamiltonian $H_{\rm s}(t)$ and coupling operator $L_{\rm s}(t)$ while the auxiliary quantum I/O system in the compound bath has the  Hamiltonian $H_{\rm a}(t)$ and coupling operator $L_{\rm a}(t)$. The unitary propagator of the cascaded system is given by 
the QSDE:
\begin{eqnarray*}
dU_t &=& \left(\vphantom{dB_t^{\dagger}}-i (H_{\rm tot}(t) + (1/2)L_{\rm sa}(t)^{\dag}L_{\rm sa}(t))dt \right.\\
&&\quad \left. + dB_t^{\dagger} L_{\rm sa}(t)- L_{\rm sa}^{\dag}(t)dB_t\right) U_t,\;U_0=I,
\end{eqnarray*}
where $H_{\rm tot}(t)= H_{\rm s}(t) + H_{\rm a}(t) + H_{\rm sa}(t)$, and $H_{\rm sa}(t) =( L_{\rm s}^{\dag}(t)L_{\rm a}(t) -L_{\rm a}^{\dag}(t) L_{\rm s}(t))/(2i)$  is the direct coupling interaction Hamiltonian between the system and auxiliary induced by the cascade connection, and $L_{\rm sa}(t) = L_{\rm s}(t) + L_{\rm a}(t)$.  
 
\subsection{General Markovian embedding}
In general, a Markovian embedding can have both direct coupling and more general instantaneous feedback interconnections (mediated by traveling quantum fields) between the principal  and auxiliary \cite{GJ08,CKS17}, with cascade connections \cite{GJ09} being a special case.  This class of model was also considered in \cite{JG10} in which the auxiliary is referred to as an exosystem. This subsection details the class of general embeddings considered in this paper,  involving multiple compound noise sources, thus multiple auxiliary systems. 

\begin{figure}
\centering
\includegraphics[scale=0.4]{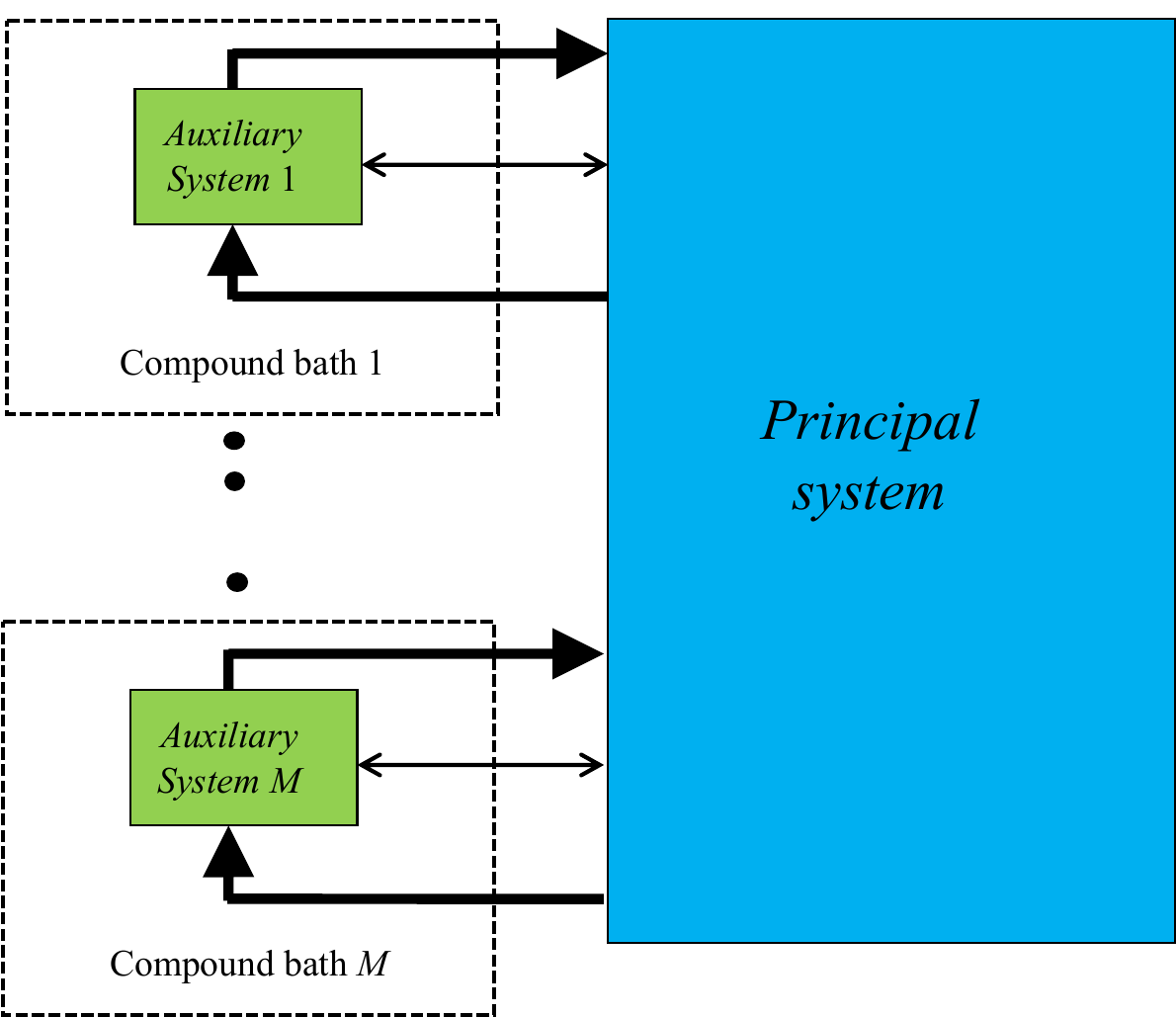}
\caption{A general Markovian embedding. The principal quantum system is coupled to $M$ compound baths. Each compound bath labeled $l$ consists of an auxiliary system that is coupled to one or more quantum fields. The principal system is coupled to each compound bath through a direct interaction term (thin bidirectional arrows) and instantaneous feedback interconnections by mediated by output quantum fields (thick one directional arrows).} \label{fig:general} 
\end{figure}

Consider a general Markovian embedding of a non-Markovian principal system coupled to $M$ compound baths consisting of an auxiliary system that is coupled to one or more quantum fields in the vacuum state, see Fig.~\ref{fig:general}. Each thin biredirectional arrow in Fig.~\ref{fig:general} denotes direct coupling  while a thick one directional arrow denotes instantaneous quantum feedback interconnections mediated by  input and output quantum fields; see \cite{GJ08,GJ09,CKS17} for details. Each thick one way arrow represents one or several  quantum fields. 

\section{Coupled non-Markovian quantum stochastic  and quantum master equations}
\label{sec:coupled-non-Markov}
This section gives a derivation of the stochastic master equation (SME) and quantum master equation (QME) for a non-Markovian system from its Markovian embedding. Both the SME (QME) take the form of coupled stochastic differentital equations/SDEs (ordinary differential equations/ODEs). This is similar to the coupled equations that have been obtained previously for (non-Markovian) principal systems driven by traveling fields in single-photon, multi-photon and more general matrix product states \cite{GJNC12,GJN13,GJN14,SZX16}. 

Consider a general Markovian embedding with M compound baths consisting of a finite-dimensional auxiliary quantum system that is coupled to one or more traveling quantum fields in the vacuum state, as shown in Fig.~\ref{fig:general}. Each auxiliary system in the compound bath labelled $l$ is coupled to $m_l$ quantum fields with annihilation and creation operators labelled as $B^{(l)}_{1k,t} $ and $B^{(l)\dag}_{1k,t}$ for  $k=1,\ldots,m_{l1}$, which couple the auxiliary and principal via a valid feedback interconnection. This interconnection is represented by a time-dependent coupling operator $L^{(l)}_{1k}(t)$ that only acts on the principal and auxiliary system $l$. The specific form of $L^{(l)}_{1k}(t)$ is determined by the topology of the interconnection, as detailed in \cite{GJ08}.
In addition, auxiliary system $l$ is also coupled to $m_{l2}$ other quantum fields with annihilation and creation operators labelled $B^{(l)}_{2k}(t)$ and $B^{(l)\dag}_{2k}(t)$ but these fields only couple to the auxiliary and not to the principal. That is, the quantum fields with subscript $2k$ couple via an operator $L^{(l)}_{2k}(t)$ that acts only on auxiliary system $l$ (and not on the principal). 

Finally, consider the case where the principal is also coupled to a single probe quantum field prepared in the vacuum state, which can be measured to retrieve information about the principal through quantum filtering \cite{BvHJ07}. Probe quantities are denoted by a superscript $(0)$, with annihilation and creation operators $B^{(0)}_t$ and $B^{(0)\dag}_t$. The coupling between the principal and the probe is given by a time-varying operator $L^{(0)}(t)$ that acts only on the principal. The joint density operator of the system and auxiliary conditional on the outcome of continuous measurement of $Y^{Q}_{o,t}$ of the probe (recall Section \ref{sec:background}), denoted by $\varrho_{\mathrm{sa},t}$, evolves according to the stochastic master equation (SME) \cite{BvHJ07}: 
\begin{eqnarray}
d\varrho_{\mathrm{sa},t} &=& \left(\vphantom{\sum_{j=1}^M} i\left[\varrho_{\mathrm{sa},t}, H_s(t) + \sum_{l=1}^{M} (H^{(l)}_{\mathrm{a}}(t) + H^{(l)}_{\mathrm{sa}}(t)) \right] \right.  \notag \\ 
&&\left. + \mathcal{D}_t^{(0)}(\varrho_{\mathrm{sa},t})+ \sum_{l=1}^M \sum_{\ell =1}^2 \mathcal{D}_t^{(l \ell)}(\varrho_{\mathrm{sa},t}) \right)dt\notag \\
&&+ \left(L^{(0)}(t)\varrho_{\mathrm{sa},t} + \varrho_{\mathrm{sa},t} L^{(0)\dag}(t)  \right. \notag \\
&&\;  \left. - \varrho_{\mathrm{sa},t} \mathrm{Tr}((L^{(0)}(t) + L^{(0) \dag}(t))\varrho_{\mathrm{sa},t})\right) dI_t,  \label{eq:sa-SME}
\end{eqnarray}
where $\mathcal{D}_t^{(0)}(\cdot) =  L^{(0)}(t) \cdot  L^{(0)}(t)^{\dag} -\frac{1}{2} \{L^{(0)}(t)^{\dag} L^{(0)}(t), \cdot\}$,
\begin{eqnarray*} 
\mathcal{D}_t^{(l\ell)}(\cdot) &=& \sum_{k=1}^{m_{l\ell}} \left( L^{(l)}_{\ell k}(t) \cdot  L^{(l)}_{\ell k}(t)^{\dag} -\frac{1}{2} \{L^{(l)}_{\ell k}(t)^{\dag} L^{(l)}_{\ell k}(t), \cdot\} \right) \\
dI_t &=& dY^Q_{o,t} -  \mathrm{Tr}(\varrho_{\mathrm{sa},t}(L^{(0)}(t) + L^{(0)\dag}(t))dt.
\end{eqnarray*}
Here $I_t$ is the innovations process of the SME, which is a Wiener process.  Let $\mathfrak{h}_{\mathrm{a}_l}$ be  the  $d_l$-dimensional Hilbert space of the $l$-th auxiliary spanned by a set of orthogonal basis vectors $\{|\phi^{(l)}_k \rangle \}_{k=1,2,\ldots,d_l}$. Then $\mathfrak{h}_{\mathrm{a}}=\otimes_{l=1}^M \mathfrak{h}_{\mathrm{a}_l}$ is spanned by the set of orthonormal basis vectors $\biggl\{|\phi^{(1)}_{i_1} \cdots \phi^{(M)}_{i_M} \rangle = |\phi^{(1)}_{i_1} \rangle \cdots |\phi^{(M)}_{i_M} \rangle \mid i_l=1,\ldots,d_l, l =1,\ldots,M \biggr\}$. 
For any linear operator $X$ on $\mathfrak{h}_{\mathrm{s}} \otimes \mathfrak{h}_{\mathrm{a}}$, define $\langle \phi^{(1)}_{j_1}   \cdots \phi^{(M)}_{j_M} | X   |\phi^{(1)}_{k_1}  \cdots \phi^{(M)}_{k_M} \rangle =(I \otimes \langle \phi^{(1)}_{j_1}  \cdots \phi^{(M)}_{j_M} |) X  (I \otimes  |\phi^{(1)}_{k_1}  \cdots \phi^{(M)}_{k_M} \rangle)$,
where $I$ is the identity operator on $\mathfrak{h}_s$. 
Recall the identity,
\begin{eqnarray*}
\lefteqn{I \otimes |\phi^{(1)}_{j_1}  \cdots \phi^{(M)}_{j_M} \rangle \langle \phi^{(1)}_{k_1}  \cdots \phi^{(M)}_{k_M} |} \\
& =(I \otimes |\phi^{(1)}_{j_1}  \cdots \phi^{(M)}_{j_M} \rangle)(I \otimes \langle \phi^{(1)}_{k_1}  \cdots \phi^{(M)}_{k_M} |)
\end{eqnarray*}
and define the principal system operator
$$
\varrho_{\mathrm{s},t}^{j_{1:M};k_{1:M}} = \langle \phi^{(1)}_{j_1}   \cdots \phi^{(M)}_{j_M} | \varrho_{\mathrm{sa},t}   |\phi^{(1)}_{k_1}  \cdots \phi^{(M)}_{k_M} \rangle
$$
on $\mathfrak{h}_{\mathrm{s}}$. Note the following identity: 
\begin{equation}
\varrho_{\mathrm{s},t} =\mathrm{Tr}_{\mathfrak{h}_{\mathrm{a}}}(\varrho_{\mathrm{sa},t})=\sum_{i_1,\ldots,i_M} \varrho_{\mathrm{s},t}^{i_{1:M};i_{1:M}}. \label{eq:varrho-s} 
\end{equation}

By inserting the resolution of identity $I =  \sum_{i_1,i_2,\ldots,i_M} |\phi^{(1)}_{i_1}  \cdots \phi^{(M)}_{i_M} \rangle  \langle \phi^{(1)}_{i_1}  \cdots \phi^{(M)}_{i_M} |$   on $\mathfrak{h}_{\mathrm{a}}$, it follows
\begin{eqnarray}
\lefteqn{\langle \phi^{(1)}_{j_1}  \cdots \phi^{(M)}_{j_M} |  H_{\mathrm{s}}(t) \varrho_{\mathrm{sa},t}   |\phi^{(1)}_{k_1}  \cdots \phi^{(M)}_{k_M} \rangle} \notag\\
&=& \langle \phi^{(1)}_{j_1}  \cdots \phi^{(M)}_{j_M} |  H_{\mathrm{s}}(t) \notag \\
&& \qquad \times \left(\sum_{i_1,i_2,\ldots,i_M} |\phi^{(1)}_{i_1}  \cdots \phi^{(M)}_{i_M} \rangle  \langle \phi^{(1)}_{i_1}  \cdots \phi^{(M)}_{i_M} | \right)  \notag \\
&&\qquad \times \varrho_{\mathrm{sa},t}   |\phi^{(1)}_{k_1}  \cdots \phi^{(M)}_{k_M} \rangle \notag \\
&=& \sum_{i_1,\ldots,i_M} \langle \phi^{(1)}_{j_1}  \cdots \phi^{(M)}_{j_M} |  H_{\mathrm{s}}(t) |\phi^{(1)}_{i_1}  \cdots \phi^{(M)}_{i_M} \rangle \notag \\
&&\quad \times \langle \phi^{(1)}_{i_1}  \cdots \phi^{(M)}_{i_M} | \varrho_{\mathrm{sa},t}   |\phi^{(1)}_{k_1}  \cdots \phi^{(M)}_{k_M} \rangle \notag \\
&=& \sum_{i_1,\ldots,i_M}   H_{\mathrm{s}}(t)   \Pi_{r} \delta_{j_r i_r}  \varrho_{\mathrm{s},t}^{i_{1:M};k_{1:M}}.
\end{eqnarray}
where $\delta_{jk}$ is the Kronecker delta and the last line uses the fact that $H_{\mathrm{s}}(t)$ acts only the principal. 
By the same process,
\begin{eqnarray*}
\lefteqn{\langle \phi^{(1)}_{j_1}  \cdots \phi^{(M)}_{j_M} |  \varrho_{\mathrm{sa},t} H_{\mathrm{s}}(t)   |\phi^{(1)}_{k_1}  \cdots \phi^{(M)}_{k_M} \rangle} \notag \\
&=& \sum_{i_1,\ldots,i_M}  \varrho_{\mathrm{sa},t}^{j_{1:M};i_{1:M}}  H_{\mathrm{s}}(t) \Pi_{r} \delta_{i_r k_r}. 
\end{eqnarray*}
Therefore,
\begin{eqnarray}
\lefteqn{\langle \phi^{(1)}_{j_1}  \cdots \phi^{(M)}_{j_M} |  [ \varrho_{\mathrm{sa},t}, H_{\mathrm{s}}(t)] |\phi^{(1)}_{k_1}  \cdots \phi^{(M)}_{k_M} \rangle} \notag \\
&=&  \sum_{i_1,\ldots,i_M} \left( \varrho_{\mathrm{s},t}^{j_{1:M};i_{1:M}} H_{\mathrm{s}}(t)  \Pi_{r} \delta_{i_r k_r} \right. \notag \\
&&\quad \left. - H_{\mathrm{s}}(t)  \varrho_{\mathrm{s},t}^{i_{1:M};k_{1:M}} \Pi_{r} \delta_{j_r i_r} \right). \label{eq:SME-1}
\end{eqnarray}
Similarly, it can be shown that 
\begin{align}
\lefteqn{\langle \phi^{(1)}_{j_1}  \cdots \phi^{(M)}_{j_M} |  [ \varrho_{\mathrm{sa},t}, H^{(l)}_{\mathrm{a}}(t) + H_{\mathrm{sa}}^{(l)}(t)]} \notag\\
\lefteqn{ |\phi^{(1)}_{k_1}  \cdots \phi^{(M)}_{k_M} \rangle} \notag\\
 &= \sum_{i_1,\ldots,i_M} \left( (\varrho_{\mathrm{s},t}^{j_{1:M};i_{1:M}} \Pi_{r,r \neq l} \delta_{i_r k_r}  \right.  \notag \\
&\quad   \times (   \langle \phi^{(l)}_{i_l} | H^{(l)}_{\mathrm{a}} | \phi^{(l)}_{k_l}\rangle  I + \langle \phi^{(l)}_{i_l} | H_{\mathrm{sa}}^{(l)}(t)| \phi^{(l)}_{k_l}\rangle) \notag \\
&\quad - ( \langle \phi^{(l)}_{j_l} |  H^{(l)}_{\mathrm{a}} | \phi^{(l)}_{i_l}\rangle I + \langle \phi^{(l)}_{j_l} | H_{\mathrm{sa}}^{(l)}(t)| \phi^{(l)}_{i_l}\rangle) \notag \\
&\quad \left.  \times \varrho_{\mathrm{s},t}^{i_{1:M};k_{1:M}} \Pi_{r, r \neq l}  \delta_{j_r i_r} )\right), \label{eq:SME-2}
\end{align} 
using the fact that $H^{(l)}_{\mathrm{a}}$ acts only on the $l$-th ancilla and $H^{(l)}_{\mathrm{sa}}$ acts only on the system and $l$-th ancilla.  Note that the identity on the right hand side is an identity on the principal. 

For the second and third terms on the right hand side of \eqref{eq:sa-SME}, recalling that $L^{(l)}_{1k}(t)$ acts on  the principal and the $l$-th auxiliary while $L^{(l)}_{2k}(t)$ acts only on auxiliary $l$, for $\ell=1,2$, 
\begin{eqnarray}
\lefteqn{\langle \phi^{(1)}_{j_1}  \cdots \phi^{(M)}_{j_M} |  \left( L^{(l)}_{\ell k}(t) \varrho_{\mathrm{sa},t}   L^{(l)\dag}_{\ell k}(t) \right.} \notag \\
\lefteqn{\left.-\frac{1}{2} \{L^{(l)\dag}_{\ell k}(t) L^{(l)}_{ \ell k}(t) , \varrho_{\mathrm{sa},t}\} \right) |\phi^{(1)}_{k_1}  \cdots \phi^{(M)}_{k_M} \rangle} \notag\\
&=& \sum_{r_1,\ldots,r_M} \sum_{s_1,\ldots,s_M}  \langle \phi^{(l)}_{j_l} | L^{(l)}_{\ell k}(t) | \phi^{(l)}_{r_l} \rangle \prod_{p, p \neq l} \delta_{j_p r_p} \notag \\
&&\quad \times \varrho_{\mathrm{s},t}^{r_{1:M};s_{1:M}}  \langle \phi^{(l)}_{s_l} | L^{(l) \dag}_{\ell k}| \phi^{(l)}_{k_l} \rangle \prod_{q, q \neq l} \delta_{s_q k_q} \notag \\
&&\quad -\frac{1}{2}\sum_{r_1,\ldots,r_M} \left(  \vphantom{ \prod_{p, p \neq l} \delta_{j_p r_p} }  \langle \phi^{(l)}_{j_l} | L^{(l)\dag}_{\ell k}(t)  L^{(l)}_{\ell k}(t) | \phi^{(l)}_{r_l}\rangle \prod_{p, p \neq l} \delta_{j_p r_p}  \right. \notag \\
&& \quad \times   \varrho_{\mathrm{s},t}^{r_{1:M};k_{1:M}} +   \varrho_{\mathrm{s},t}^{j_{1:M};r_{1:M}}  \prod_{p, p \neq l} \delta_{r_p k_p} \notag\\
&&\quad \left.  \times \langle \phi^{(l)}_{r_l} | L^{(l)\dag}_{\ell k}(t)  L^{(l)}_{\ell k}(t) | \phi^{(l)}_{k_l}\rangle   \right). \label{eq:SME-3}
\end{eqnarray} 

By an analogous computation for the final term of \eqref{eq:sa-SME}, since $L^{(0)}(t)$ only acts on the principal system, we find that 
\begin{eqnarray}
\lefteqn{\langle \phi^{(1)}_{j_1}  \cdots \phi^{(M)}_{j_M} |   (L^{(0)}(t) \varrho_{\mathrm{sa},t}  +   \varrho_{\mathrm{sa},t} L^{(0)\dag}(t))  |\phi^{(1)}_{k_1}  \cdots \phi^{(M)}_{k_M} \rangle } \notag \\
&=& \sum_{r_1,\ldots,r_M} \left( L^{(0)}(t) \varrho_{\mathrm{s},t}^{r_{1:M};k_{1:M}}  \Pi_{p} \delta_{j_p r_p} \right. \notag\\
&&\quad \left. +  \varrho_{\mathrm{s},t}^{j_{1:M};r_{1:M}} L^{(0)\dag}(t) \Pi_{p} \delta_{r_p k_p} \right)  \label{eq:SME-4} \\
\lefteqn{ \langle \phi^{(1)}_{j_1}  \cdots \phi^{(M)}_{j_M} | \varrho_{\mathrm{sa},t} \mathrm{Tr}((L^{(0)}(t) + L^{(0) \dag}(t))\varrho_{\mathrm{sa},t})} \notag\\
\lefteqn{ \quad |\phi^{(1)}_{k_1}  \cdots \phi^{(M)}_{k_M} \rangle} \notag \\
&=&  \varrho_{\mathrm{s},t}^{j_{1:M};k_{1:M}}  \notag\\
&&\; \times\sum_{i_1,\ldots,i_M} \mathrm{Tr}((L^{(0)}(t) + L^{(0) \dag}(t)) \varrho_{\mathrm{s},t}^{i_{1:M};i_{1:M}}). \label{eq:SME-5}
\end{eqnarray} 
where the last line uses \eqref{eq:varrho-s}. Finally, by sandwiching both sides of \eqref{eq:sa-SME} between $\langle \phi^{(1)}_{j_1}  \cdots \phi^{(M)}_{j_M} |$ and  $|\phi^{(1)}_{k_1}  \cdots \phi^{(M)}_{k_M} \rangle$ and inserting \eqref{eq:SME-1} to \eqref{eq:SME-5} to the right hand,  one then obtains an operator-valued SDE for $\varrho^{j_{1:M},k_{1:M}}_{s,t}$ as given in \eqref{eq:coupled-SME}. 
\begin{strip}
\begin{eqnarray}
d\varrho_{\mathrm{s},t}^{j_{1:M};k_{1:M}} &=&\left( i \mathscr{H}_t^{j_{1:M};k_{1:M}}(\varrho_{\mathrm{s},t})  + \mathscr{D}_t^{j_{1:M};k_{1:M}}(\varrho_{\mathrm{s},t})\ \right)dt + \mathscr{B}_{0,t}^{j_{1:M};k_{1:M}}(\varrho_{\mathrm{s},t}) dI_t, \label{eq:coupled-SME}\\
\mathscr{H}_t^{j_{1:M};k_{1:M}}(\varrho_{\mathrm{s},t}) &=& \mathscr{H}_{\mathrm{s},t}^{j_{1:M};k_{1:M}}(\varrho_{\mathrm{s},t})   +  \sum_{l=1}^M   \mathscr{H}_{l,t}^{j_{1:M};k_{1:M}}(\varrho_{\mathrm{s},t}), \label{eq:H-coeffs} \\
\mathscr{D}_t^{j_{1:M};k_{1:M}}(\varrho_{\mathrm{s},t}) &=&  \mathcal{D}_t^{(0)}(\varrho_{\mathrm{s},t}^{j_{1:M};k_{1:M}}) + \sum_{l=1}^M \sum_{\ell=1}^2 \sum_{k=1}^{m_{l\ell}} \mathscr{D}_{l\ell k,t}^{j_{1:M};k_{1:M}}(\varrho_{\mathrm{s},t}), \label{eq:D-coeffs}
\end{eqnarray}
\end{strip} 

In  \eqref{eq:coupled-SME}-\eqref{eq:D-coeffs}, the maps (superoperators)  $\mathscr{H}_{\mathrm{s},t}^{j_{1:M};k_{1:M}}$, $\mathscr{H}_{l,t}^{j_{1:M};k_{1:M}}$, $\mathscr{D}_{l \ell k,t}^{j_{1:M};k_{1:M}}$ and  $\mathscr{B}_{0,t}^{j_{1:M};k_{1:M}}$ are defined as follows: 
\begin{itemize}
\item $\mathscr{H}_{\mathrm{s},t}^{j_{1:M};k_{1:M}}(\cdot)$ is defined by the right hand side of \eqref{eq:SME-1} 

\item $\mathscr{H}_{l,t}^{j_{1:M};k_{1:M}}(\cdot)$ is defined by the right hand side of \eqref{eq:SME-2}

\item $\mathscr{D}_{l \ell k,t}^{j_{1:M};k_{1:M}}(\cdot)$ is defined by the right hand side of \eqref{eq:SME-3}

\item $\mathscr{B}_{0,t}^{j_{1:M};k_{1:M}}(\cdot)$ is defined by subtracting  the right hand side of \eqref{eq:SME-5} from the right hand side of \eqref{eq:SME-4}. 
\end{itemize}
The conditional density operator $\varrho_{\mathrm{s},t} $ of the principal system alone is then given by the summation in \eqref{eq:varrho-s}.  

Since the innovations process $I_t$ is a martingale with respect to the $\sigma$-algebra generated by $Y^Q_{0:s}$, $s \leq t$ with zero mean for each $t$, the expectation of the term containing $dI_t$ on the right hand side of \eqref{eq:coupled-SME} vanishes. Therefore, by taking expectation on both sides of \eqref{eq:coupled-SME}, $\rho_{\mathrm{s},t}^{i_{1:M};j_{1:M}}$ satisfies the coupled operator-valued differential equation (the analogue of the Lindblad quantum master equation) given by  \eqref{eq:coupled-ME} below. We then get that  $\rho_{\mathrm{s},t} = \sum_{i_1,\ldots,i_M} \rho_{\mathrm{s},t}^{i_{1:M};i_{1:M}}$. 

\begin{align}
\dot{\rho}_{\mathrm{s},t}^{j_{1:M};k_{1:M}}= i \mathscr{H}_t^{j_{1:M};k_{1:M}}(\rho_{\mathrm{s},t})   + \mathscr{D}_t^{j_{1:M};k_{1:M}}(\rho_{\mathrm{s},t}).\label{eq:coupled-ME}
\end{align}

Although the derivation above is given for a continuous measurement of $Y^Q_{o,t}$, the calculations for the non-Markovian SME can be straightforwardly modified for the phase quadrature measurement of $Y^P_{o,t}$ and photon counting measurements of $\Lambda_{o,t}$ on the probe. In the case of photon counting measurements,  the stochastic term on the right hand side of \eqref{eq:sa-SME} has to be modified accordingly, see, e.g., \cite{BvHJ07,GJNC12,GJN13,GJN14}.  However, whatever continuous measurement is performed the coupled quantum master equation remains the same,  as given by \eqref{eq:coupled-ME}. 
\begin{remark}
Observe that \eqref{eq:coupled-SME} and \eqref{eq:coupled-ME} also hold for a purely Hamiltonian coupling between the principal and auxiliaries,  without any white noise fields ($\mathscr{D}_t^{j_{1:M};k_{1:M}}(\varrho_{\mathrm{s},t})=0$). The setting of this paper is quite general.
\end{remark}




\section{Conclusion}
\label{sec:conclu}
This work has investigated non-Markovian quantum systems that have a Markovian embedding. This embedding consists of the principal non-Markovian quantum system of interest that is coupled to one or more compound baths, which each consists of an auxiliary quantum I/O system coupled to traveling quantum fields. The coupling between the principal and the auxiliary can be
facilitated by a direct interaction between the two quantum systems as well as general instantaneous feedback interconnections facilitated by the quantum fields. Starting from this Markovian embedding, a set of coupled quantum stochastic master equations and quantum master equations were derived for the non-Markovian quantum system. 

Future work that can follow from the results herein include  open-loop and feedback control of continuous-time non-Markovian systems and model reduction of such systems, and studying quantum networks with delayed feedback \cite{NG17} under continuous observation.  The results may also provide additional insights on the general structure of the evolution of continuous-time non-Markovian quantum systems. 



\bibliographystyle{ieeetran}
\bibliography{cdc2023}



\end{document}